\documentclass[a4paper,11pt]{article}
\usepackage{amsmath}
\usepackage{ragged2e} 
\usepackage[utf8]{inputenc}
\usepackage{cite}
\usepackage{authblk}
\usepackage{booktabs}
\usepackage{graphicx} % figuras
\usepackage{subfigure} % subfiguras
\usepackage{slashed}\usepackage{float}
\usepackage{amsmath,amssymb,lmodern,setspace}
\usepackage{multirow} 
\usepackage{xcolor}
\usepackage[capitalise]{cleveref}
\usepackage{soul}

\title{$\tau$ data-driven evaluation of Euclidean windows for the hadronic vacuum polarization}

\author[1,2]{Pere Masjuan~\footnote{masjuan@ifae.es}}
\author[2]{Alejandro Miranda~\footnote{jmiranda@ifae.es}}
\author[3]{Pablo Roig~\footnote{pablo.roig@cinvestav.mx}}

\affil[1]{Grup de F\'isica Te\`orica, Departament de F\'isica\\
Universitat Aut\`onoma de Barcelona, 08193 Bellaterra (Barcelona), Spain.}
\affil[2]{Institut de F\'isica d'Altes Energies (IFAE) and
The Barcelona Institute of Science and Technology (BIST),
Campus UAB, 08193 Bellaterra (Barcelona), Spain.}
\affil[3]{Departamento de F\'isica, Centro de Investigaci\'on y de Estudios Avanzados del Instituto Polit\'ecnico Nacional\\
Apdo. Postal 14-740,07000 Ciudad de M\'exico, M\'exico}

\date{}

\begin{document}
\maketitle
\abstract{We compute for the first time the $\tau$ data-driven Euclidean windows for the hadronic vacuum polarization contribution to the muon $g-2$. We show that $\tau$-based results agree with the available lattice window evaluations and with the full result. On the intermediate window, where all lattice evaluations are rather precise and agree, $\tau$-based results are compatible with them. This is particularly interesting, given that the disagreement of the $e^+e^-$ data-driven result with the lattice values in this window is the main cause for their discrepancy, affecting the interpretation of the $a_\mu$ measurement in terms of possible new physics.}

\section{Introduction}

The first Fermilab measurement of the muon anomalous magnetic moment ($a_\mu=(g_\mu-2)/2$, with $g$ the gyromagnetic factor) confirmed \cite{Muong-2:2021ojo} the final result from Brookhaven \cite{Muong-2:2006rrc}, yielding the world average 
\begin{equation}\label{eq:amuexp}
a_\mu^{\mathrm{Exp}}\times10^{11}=116592061(41)\,.
\end{equation}
This reaffirmed and strengthened the interest of the high-energy physics community in this observable, given its $4.2\sigma$ tension with the Standard Model prediction \cite{Davier:2017zfy, Keshavarzi:2018mgv, Colangelo:2018mtw, Hoferichter:2019mqg, Davier:2019can, Keshavarzi:2019abf, Kurz:2014wya,FermilabLattice:2017wgj,Budapest-Marseille-Wuppertal:2017okr, RBC:2018dos,Giusti:2019xct,Shintani:2019wai,FermilabLattice:2019ugu,Gerardin:2019rua,Aubin:2019usy,Giusti:2019hkz,Melnikov:2003xd,Masjuan:2017tvw,Colangelo:2017fiz,Hoferichter:2018kwz,Gerardin:2019vio,Bijnens:2019ghy,Colangelo:2019uex,Pauk:2014rta,Danilkin:2016hnh,Jegerlehner:2017gek,Knecht:2018sci,Eichmann:2019bqf,Roig:2019reh,Colangelo:2014qya,Blum:2019ugy,Aoyama:2012wk,Aoyama:2019ryr,Czarnecki:2002nt,Gnendiger:2013pva}, obtained within the Muon g-2 Theory Initiative \cite{Aoyama:2020ynm}~\footnote{See later developments in e.g. Refs.~\cite{Colangelo:2022jxc,Bijnens:2020xnl,Chao:2020kwq,Colangelo:2020lcg,Hoid:2020xjs,Masjuan:2020jsf,Ludtke:2020moa,Bijnens:2021jqo,Cappiello:2021vzi,Chao:2021tvp,Colangelo:2021moe,Colangelo:2021nkr,Danilkin:2021icn,Hoferichter:2021wyj,Leutgeb:2021mpu,Miramontes:2021exi,Zanke:2021wiq,Biloshytskyi:2022ets,Colangelo:2022prz,Bijnens:2022itw,Boito:2022rkw,Boito:2022dry,Stamen:2022uqh,Blum:2023qou,Ludtke:2023hvz, Davier:2023hhn, Wang:2023njt}.}
\begin{equation}
    a_\mu^{\mathrm{SM}}\times10^{11}=116591810(43)\,.\label{eq:amuSM}
\end{equation}
However, the BMWc  lattice Collaboration published \cite{Borsanyi:2020mff} a precise lattice-QCD evaluation of the hadronic vacuum polarization (HVP) contribution with the result
\begin{equation}
  a_\mu^{\mathrm{BMWc}}\times10^{10}= 707.5\pm5.5\,,
\end{equation}
which yields an $a_\mu$ evaluation at $1.5\sigma$ from $a_\mu^{\mathrm{Exp}}$, Eq.~(\ref{eq:amuexp}), and at $2.1\sigma$ from the data-driven result, based on $\sigma(e^+e^-\to$hadrons$)$, that was employed to get Eq.~(\ref{eq:amuSM}). A direct comparison between the BMWc and data-driven predictions for the HVP contribution can be found in Table~\ref{HVP:tab4.4c}.

Recently, other three accurate lattice evaluations, by the Mainz CLS \cite{Ce:2022kxy}, the ETMC   \cite{ExtendedTwistedMass:2022jpw} and the RBC/UKQCD \cite{Blum:2023qou} collaborations, have agreed within the so-called intermediate window (defined below, in Eq.~(\ref{eq:weight_function})) with the BMWc result, having commensurate uncertainties.

Reference \cite{Colangelo:2022vok}, which sets the foundations to start this project, has translated the data-driven results to the three different windows in Euclidean time used by lattice practitioners, to scrutinize the root of the current discrepancy between both groups of results. This is a crucial endeavour, since it currently limits the implications on new physics of the $a_\mu$ measurements, as may help to pinpoint in what energy domain discrepancies emerge~\footnote{Aditionally, a data-driven estimation for the isospin-limit light-quark connected component of the intermediate-window contribution ($a_\mu^{\text{win,lqc}}$)~\cite{Benton:2023dci} is in significant tension with recent very-precise lattice-QCD predictions~\cite{Borsanyi:2020mff,Lehner:2020crt,Wang:2022lkq,Aubin:2022hgm,Ce:2022kxy,ExtendedTwistedMass:2022jpw,FermilabLatticeHPQCD:2023jof,Blum:2023qou}, showing that the difference between data-driven and lattice-QCD outcomes for $a_\mu^{\text{win}}$ is almost completely ascribed to the light-quark connected contribution.}. It is also a timely task, as we expect the soon release of the second $a_\mu$ FNAL value, with improved precision, in the forthcoming months~\footnote{Indeed it was during the refereeing of this article, \cite{Muong-2:2023cdq}.}.

The situation has become even more puzzling with the recent accurate $\sigma(e^+e^-\to\pi^+\pi^-)$~\footnote{This contribution yields a bit more than $70\%$ of $a_\mu^{\text{HVP,LO}}$.} CMD-3 measurement \cite{CMD-3:2023alj,CMD-3:2023rfe}, which would -on its own- reduce the discrepancy with $a_\mu^{\mathrm{Exp}}$ to less than two standard deviations. Consequently, it is in tension with the previous measurements of this reaction, by CMD-2 \cite{CMD-2:2003gqi,CMD-2:2005mvb,Aulchenko:2006dxz,CMD-2:2006gxt}, SND \cite{Achasov:2006vp,SND:2020nwa}, KLOE \cite{KLOE:2008fmq,KLOE:2010qei,KLOE:2012anl,KLOE-2:2017fda}, BaBar \cite{BaBar:2012bdw}, BES \cite{BESIII:2015equ} and CLEO \cite{CLEO:2005tiu}. Dedicated studies related to the radiative corrections employed in the Monte Carlo generators used by these experiments \cite{WorkingGrouponRadiativeCorrections:2010bjp} (see e.g. Refs.~\cite{Campanario:2019mjh,Ignatov:2022iou,Colangelo:2022lzg}) could provide more insight into this matter.

Within the data-driven evaluations of $a_\mu^{\text{HVP}}$, $\tau^-\to\nu_\tau$hadrons was proposed in 1997 to reduce the error (by $\sim37\%$ then) of the method  when using only $e^+e^-\to$hadrons in LEP times \cite{Alemany:1997tn}. Through the years, this alternative data-driven method has always been \cite{Cirigliano:2001er, Cirigliano:2002pv,Davier:2010fmf,Davier:2010nc,Davier:2013sfa,Miranda:2020wdg} approximately  $[2,2.5]\sigma$ away from the $a_\mu^{\mathrm{Exp}}$ world average, a situation which seems now favored by the lattice QCD evaluations of $a_\mu^{\text{HVP}}$~\footnote{The difference between both data-driven groups of results could also be due to non-standard interactions modifying slightly -but noticeably at this precision- the di-pion $\tau$ decays \cite{Miranda:2018cpf,Cirigliano:2018dyk,Gonzalez-Solis:2020jlh,Cirigliano:2021yto}.}. Still, a direct comparison between lattice QCD results and $\tau$ data-driven methodology is not straightforward either. This motivates us to compute in this work Euclidean windows for $a_\mu^{\text{HVP}}$ using $\tau$ data and compare results. We hope our outcomes may shed some light on this puzzling situation ~\footnote{We acknowledge that if good agreement was found amid $e^+e^-\to\pi^+\pi^-$ measurements, di-pion tau decays could not possibly contribute, since the error associated to the required isospin-breaking corrections would be larger than the uncertainties of the cross-section measurements.} and be useful for the lattice effort \cite{Bruno:2018ono} to compute the required isospin-breaking corrections (see below Eq.~(\ref{eq:weight_function})) entering the $\tau$-based method. In section \ref{sec:Results} we explain the needed formulae and apply them to obtain the results, which are summarized in section \ref{sec:Concl}.

\section{Results} \label{sec:Results}
The HVP contribution at leading order (LO) %, $\mathcal{O}(\alpha^2)$, 
in the data-driven approach is given by~\cite{Bouchiat:1961lbg,Durand:1962zzb,Brodsky:1967sr,Gourdin:1969dm}  
%%%%%%%%%%%%%%%%%%%%%%%%%%%%%%%%%%%%%%%%%%%%%%%%%%%%%%%%%%%%%%%%%
\begin{equation}\label{amu_dispersive}
    a_\mu^{\text{HVP,LO}}=\frac{1}{4\pi^3}\int_{s_{\text{thr}}}^{\infty}ds \,K(s)\,\sigma_{e^+e^-\to\text{hadrons}(+\gamma)}^{0}(s),
\end{equation}

%%%%%%%%%%%%%%%%%%%%%%%%%%%%%%%%%%%%%%%%%%%%%%%%%%%%%%%%%%%%%%%%%
\noindent
where $\sigma^{0}_{e^+e^-\to \text{hadrons}(+\gamma)}(s)$ is the bare hadronic cross-section, which excludes effects from vacuum polarization (VP)~\cite{Eidelman:1995ny}, and $K(s)$ is a smooth kernel concentrated at low energies~\cite{Brodsky:1967sr,Lautrup:1968tdb}
%%%%%%%%%%%%%%%%%%%%%%%%%%%%%%%%%%%%%%%%%%%%%%%%%%%%%%%%%%%%%%%%%
\begin{equation}\begin{split}
    &K(s)=\frac{x^2}{2}(2-x^2)+\frac{(1+x^2)(1+x)^2}{x^2}\left(\log(1+x)-x+\frac{x^2}{2}\right)+\frac{(1+x)}{(1-x)}x^2\log(x),
\end{split}\end{equation}
%%%%%%%%%%%%%%%%%%%%%%%%%%%%%%%%%%%%%%%%%%%%%%%%%%%%%%%%%%%%%%%%%
which is written in terms of the variable $x=\frac{1-\beta_\mu}{1+\beta_\mu}$, $\beta_\mu=\sqrt{1-4m_\mu^2/s}$. This dispersive approach revolves around the handiness of $e^+e^-$ hadronic cross-section measurements at energies below a few GeV. However, as was pointed  out early by Alemany, Davier, and H\"ocker~\cite{Alemany:1997tn}, it is also possible to use hadronic $\tau$ decays to evaluate the HVP, LO contributions to $a_\mu$. For some time this approach was competitive with the $e^+e^-$ data, although this is generally not considered the case at the moment \cite{Aoyama:2020ynm}.

Recently, a new $\tau$ data-driven approach was performed in Ref. \cite{Miranda:2020wdg}. In this section, we utilize their results to evaluate the leading HVP contribution to the anomalous magnetic moment of the muon in the so-called window quantities~\cite{RBC:2018dos,Lehner:2020crt}. For this enterprise, we make use of the weight functions in center-of-mass energy $\tilde{\Theta}(s)$  from Eq. (12) in Ref. \cite{Colangelo:2022vok} which are related to those in Euclidean time~\cite{RBC:2018dos}
%%%%%%%%%%%%%%%%%%%%%%%%%%%%%%%%%%%%%%%%%%%%%%%%%%%%%%%%%%%%%%%%%
\begin{equation}\label{eq:weight_function}\begin{split}
    \Theta_{SD}(t)&=1-\Theta(t,t_0,\Delta),\\
    \Theta_{win}(t)&=\Theta(t,t_0,\Delta)-\Theta(t,t_1,\Delta),\\
    \Theta_{LD}(t)&=\Theta(t,t_1,\Delta),\\
    \Theta(t,t^\prime,\Delta)&=\frac{1}{2}\left(1+\tanh{\frac{t-t^\prime}{\Delta}}\right).
\end{split}\end{equation}
%%%%%%%%%%%%%%%%%%%%%%%%%%%%%%%%%%%%%%%%%%%%%%%%%%%%%%%%%%%%%%%%%
The subscripts in Eq. (\ref{eq:weight_function}) refer to the short-distance ($SD$), intermediate ($win$, although we will use $int$ in the following) and long-distance ($LD$) contributions with parameters 

\begin{equation}\label{eq:win_parameters}
    t_0=0.4\text{ fm}, \quad t_1=1.0\text{ fm},\quad \Delta=0.15\text{ fm},
\end{equation}
which correspond to inverse energies of the order of $500,200$ and $1300$ MeV, respectively.

In what follows, we will focus on  the dominant $2\pi$ contribution only. Including isospin breaking (IB) corrections, i.e., $\mathcal{O}[(m_u-m_d)p^2]$ and $\mathcal{O}(e^2 p^2)$  contributions, the {\it bare} hadronic $e^+e^-$ cross-section  $\sigma_{\pi\pi(\gamma)}^0$ 
is related to the {\it observed} differential $\tau$ decay rate   $d\Gamma_{\pi\pi[\gamma]}$ through~\cite{Cirigliano:2001er,Cirigliano:2002pv}
%%%%%%%%%%%%%%%%%%%%%%%%%%%%%%%%%%%%%%%%%%%%%%%%%%%%%%%%%%%%%%%%%
\begin{equation}\label{eq:pipi_cross_section}
\sigma^0_{\pi\pi(\gamma)}=\left[\frac{K_\sigma(s)}{K_\Gamma(s)}\frac{d\Gamma_{\pi\pi[\gamma]}}{ds}\right]\times\frac{R_{\text{IB}}(s)}{S_{\text{EW}}},
\end{equation}
%%%%%%%%%%%%%%%%%%%%%%%%%%%%%%%%%%%%%%%%%%%%%%%%%%%%%%%%%%%%%%%%%
where
%%%%%%%%%%%%%%%%%%%%%%%%%%%%%%%%%%%%%%%%%%%%%%%%%%%%%%%%%%%%%%%%%
\begin{equation}\begin{split}
    K_\Gamma(s)&=\frac{G_F^2\vert V_{ud}\vert^2 m_\tau^3}{384\pi^3}\left(1-\frac{s}{m_\tau^2}\right)^2\left(1+\frac{2s}{m_\tau^2}\right),\\
    K_{\sigma}(s)&=\frac{\pi \alpha^2}{3s},
\end{split}\end{equation}
%%%%%%%%%%%%%%%%%%%%%%%%%%%%%%%%%%%%%%%%%%%%%%%%%%%%%%%%%%%%%%%%%
and the IB corrections are encoded in 
%%%%%%%%%%%%%%%%%%%%%%%%%%%%%%%%%%%%%%%%%%%%%%%%%%%%%%%%%%%%%%%%%
\begin{equation}
    R_{\text{IB}}(s)=\frac{\text{FSR}(s)}{G_{\text{EM}}(s)}\frac{\beta^3_{\pi^{+}\pi^{-}}}{\beta^3_{\pi^+\pi^0}}\left\vert\frac{F_V(s)}{f_{+}(s)}\right\vert^2.
\end{equation}
%%%%%%%%%%%%%%%%%%%%%%%%%%%%%%%%%%%%%%%%%%%%%%%%%%%%%%%%%%%%%%%%%
The $S_{\text{EW}}$ term in Eq. (\ref{eq:pipi_cross_section}) includes the short-distance electroweak corrections~\cite{Sirlin:1974ni,Sirlin:1977sv,Sirlin:1981ie,Marciano:1985pd,Marciano:1988vm,Marciano:1993sh,Braaten:1990ef,Erler:2002mv}. FSR refers to the Final-State-Radiation corrections to the $\pi^+\pi^-$ channel~\cite{Schwinger:1989ix,Drees:1990te}~\footnote{To our knowledge, FSR was not included in $R_{\text{IB}}$ before Ref.~\cite{Davier:2010fmf}.}, while the $G_{\text{EM}}(s)$ factor includes the QED corrections to the $\tau^-\to\pi^-\pi^0\nu_\tau$ decay with virtual plus real photon radiation. $\beta^{3}_{\pi^-\pi^+}/\beta^{3}_{\pi^-\pi^0}$ is a phase space correction owing to the $\pi^\pm - \pi^0$ mass difference. The last term in $R_{\text{IB}}(s)$ corresponds to the ratio between the neutral ($F_V(s)$) and the charged ($f_{+}(s)$) pion form factor, which includes one of the leading IB effects, the $\rho^0-\omega$ mixing correction.

The IB corrections to $a_\mu^\text{HVP, LO}$ using $\tau$ data in the dominant $\pi\pi$ channel~\cite{GomezDumm:2013sib,Gonzalez-Solis:2019iod} can be evaluated using the following expression~\cite{Davier:2010fmf}
%%%%%%%%%%%%%%%%%%%%%%%%%%%%%%%%%%%%%%%%%%%%%%%%%%%%%%%%%%%%%%%%%
\begin{equation}
    \Delta a_\mu^{\text{HVP, LO}}[\pi\pi,\tau]=\frac{1}{4\pi^3}\int_{4m_\pi^2}^{m_\tau^2}ds\, K(s)\left[\frac{K_\sigma(s)}{K_\Gamma(s)}\frac{d\Gamma_{\pi\pi[\gamma]}}{ds}\right]\left(\frac{R_{\text{IB}}(s)}{S_{\text{EW}}}-1\right),
\end{equation}
%%%%%%%%%%%%%%%%%%%%%%%%%%%%%%%%%%%%%%%%%%%%%%%%%%%%%%%%%%%%%%%%%
which measures the difference between the correct expression for $\sigma_{\pi\pi(\gamma)}^{0}$ and the naive Conserved Vector Current approximation, with $S_{\text{EW}}=1$ and $R_{\text{IB}}(s)=1$.

These contributions are summarized in Table \ref{tab:Delta_amu} for each IB correction.
\begin{itemize}
    \item The $S_{\text{EW}}$ factor ($S_{\text{EW}}=1.0233(3)$, at the scale $m_\tau$) contributes $3.0\%$, $28.7\%$ and $68.3\%$ to the complete $\Delta a_\mu^{\text{HVP, LO}}=-11.96(0)\cdot 10^{-10}$ correction for the $SD$, $int$ and $LD$ contributions, respectively~\footnote{We are correcting here our double-counting \cite{Miranda:2020wdg} of the subleading
non-logarithmic short-distance correction for quarks, as noted in ref.~\cite{Davier:2023fpl} (see ref.~\cite{Cirigliano:2023fnz}).}.
    \item The phase space (PS) correction yields $1.7\%$, $18.7\%$ and $79.6\%$ of a total of $\Delta a_\mu^{\text{HVP, LO}}=-7.47(0)\cdot 10^{-10}$ for the $SD$, $int$ and $LD$ contributions, respectively.
    \item The final state radiation (FSR) induces $2.9\%$, $27.0\%$ and $70.1\%$ of a total of $\Delta a_\mu^{\text{HVP, LO}}=+4.56(45)\cdot 10^{-10}$ for the $SD$, $int$ and $LD$ contributions, respectively.
    \item $G_{\text{EM}}(s)$ was originally computed in Ref. \cite{Cirigliano:2002pv} including those operators yielding resonance saturation of the $\mathcal{O}(p^4)$ chiral couplings in the frame of Resonance Chiral Theory (R$\chi $T)~\cite{Ecker:1988te,Ecker:1989yg,Cirigliano:2006hb,Kampf:2011ty}~\footnote{We will shorten this to `$\mathcal{O}(p^4)$' for brevity, and similarly at the next chiral order. In the first case, short-distance QCD behaviour is ensured for two point Green functions, while this is extended up to 3-point correlators in the latter.}~\footnote{Pseudoscalar pole (axial) contributions to the hadronic light-by-light piece of $a_\mu$ have been computed within R$\chi$T \cite{Roig:2014uja,Guevara:2018rhj} (\cite{Roig:2019reh}), as well as di-meson contributions to $a_\mu^{\text{HVP}}$ \cite{Qin:2020udp,Wang:2023njt}.}, which comprises Goldstone bosons and resonance fields extending the $\chi$PT framework~\cite{Gasser:1983yg,Gasser:1984gg,Leutwyler:1993iq} to higher energies. A recalculation of $G_{\text{EM}}(s)$ was performed later in \cite{Flores-Baez:2006yiq,Flores-Tlalpa:2006snz} using a Vector Meson Dominance (VMD) model~\cite{Sakurai:1960ju}.\\
    In Ref. \cite{Miranda:2020wdg} (see also \cite{Escribano:2023seb}), two of us explored the impact of the R$\chi$T operators contributing to resonance saturation at the next chiral order ($\mathcal{O}(p^6)$) on the $G_{\text{EM}}(s)$, as well as estimating the uncertainty of the original computation in \cite{Cirigliano:2002pv}. These results \cite{Miranda:2020wdg} are consistent with the earlier R$\chi$T and the VMD estimates \cite{Miranda:2020wdg}. Availing of these results%\textbf{(I think that uncertainties for the following numbers should be given)}
    , $G_{\text{EM}}(s)$ produces a correction of $\sim-3.5\%$, $\sim-17.1\%$ and $\left(-17.1\left(^{+6.1}_{-14.8}\right)\right)$  $\sim+120.6\%$ of $\Delta a_\mu^{\text{HVP, LO}}=-1.71(^{0.61}_{1.48})\cdot 10^{-10}$ at $\mathcal{O}(p^4)$, and $\sim0.4\%$, $\sim7.6\%$ and $\sim92.0\%$ of $\Delta a_\mu^{\text{HVP, LO}}=-7.59(^{6.50}_{4.56})\cdot 10^{-10}$ at $\mathcal{O}(p^6)$ for the $SD$, $int$ and $LD$ contributions, respectively. Interestingly, the $SD$ and $int$ contributions at $\mathcal{O}(p^4)$ change in sign while this is not the case at $\mathcal{O}(p^6)$ where all the contributions are always negative.
    \item The ratio of the form factors (FF) gives an overall correction of $\Delta a_\mu^{\text{HVP, LO}}=+7.13(1.48)(^{1.59}_{1.54})(^{0.85}_{0.80})\cdot 10^{-10}$, from which $\sim2.2\%$, $\sim26.8\%$ and $\sim71.0\%$ stand for the SD, int and LD corrections, respectively%\textbf{(I would again give uncertainties on these percentages)}
    . \\
    The errors quoted in this contribution correspond to the electromagnetic shifts in the widths and masses of the $\rho$ meson, and to the $\rho^0 -\omega$ mixing parameter~\cite{Urech:1995ry,Cirigliano:2002pv} (see Eqs. (5.5) and (5.6) in Ref.~\cite{Cirigliano:2002pv}), respectively~\footnote{We note that, according to refs.~\cite{Colangelo:2018mtw,Sanchez-Puertas:2021eqj,Colangelo:2022prz}, this correction is smaller in data-driven methods than in phenomenological estimates. The associated uncertainties, however, are subleading in our analysis, as we have checked in detail, accounting for these errors, in the comparisons shown in fig.~\ref{fig:KLEO_BABAR}.}. For this analysis, we use the same numerical inputs as in \cite{Miranda:2020wdg}. The central value reported in Table \ref{tab:Delta_amu} corresponds to the weighted average between the FF1 and FF2 sets~\footnote{The main distinction between FF1 and FF2 comes from the width difference between the $\rho^\pm$ and $\rho^0$ mesons ($\Delta \Gamma_\rho=\Gamma_{\rho^0}-\Gamma_{\rho^\pm}$), while the mass difference ($\Delta M_\rho=M_{\rho^0}-M_{\rho^\pm}$) and the $\theta_{\omega\rho}$ parameter are the same in both. }.
\end{itemize}

The overall correction is also consistent with those in Refs.~\cite{Cirigliano:2002pv,Davier:2010fmf,Miranda:2020wdg}.

\begin{table}[htbp]
\centering
\resizebox{14cm}{!}{\begin{tabular}{|c|cc|cc|cc|cc|}
\hline
 \multicolumn{9}{ |c| }{$\Delta a_\mu^{\text{HVP,LO}}$} \\ [0.4ex]
\hline
 & \multicolumn{2}{ |c| }{SD} & \multicolumn{2}{ |c| }{int} & \multicolumn{2}{ |c| }{LD} & \multicolumn{2}{ |c| }{Total} \\ [0.3ex]
  & $\mathcal{O}(p^4)$ & $\mathcal{O}(p^6)$ & $\mathcal{O}(p^4)$ & $\mathcal{O}(p^6)$ & $\mathcal{O}(p^4)$ & $\mathcal{O}(p^6)$ & $\mathcal{O}(p^4)$ & $\mathcal{O}(p^6)$ \\ [0.4ex]
\hline
$S_{\text{EW}}$ & \multicolumn{2}{ |c| }{$-0.36(0)$} & \multicolumn{2}{ |c| }{$-3.43(0)$} & \multicolumn{2}{ |c| }{$-8.17(0)$} & \multicolumn{2}{ |c| }{$-11.96(0)$} \\ [0.3ex]
PS & \multicolumn{2}{ |c| }{$-0.13(0)$} & \multicolumn{2}{ |c| }{$-1.39(0)$} & \multicolumn{2}{ |c| }{$-5.95(0)$} & \multicolumn{2}{ |c| }{$-7.47(0)$} \\ [0.3ex]
FSR & \multicolumn{2}{ |c| }{$+0.13(1)$} & \multicolumn{2}{ |c| }{$+1.23(12)$} & \multicolumn{2}{ |c| }{$+3.20(32)$} & \multicolumn{2}{ |c| }{$+4.56(46)$} \\ [0.3ex]
$G_{\text{EM}}$ & $+0.06(^{0}_{2})$ & $-0.03(^{9}_{7})$ & $+0.29(^{6}_{19})$ & $-0.58(^{93}_{71})$ & $-2.06(^{0.55}_{1.27})$ & $-7.00(^{5.48}_{3.78})$ & $-1.71(^{0.61}_{1.48})$ & $-7.61(^{6.50}_{4.56})$\\ [0.3ex]
FF & \multicolumn{2}{ |c| }{$+0.16(5)(1)(^{2}_{1})$} & \multicolumn{2}{ |c| }{$+1.91(49)(^{29}_{27})(^{21}_{20})$} & \multicolumn{2}{ |c| }{$+5.06(94)(^{1.29}_{1.26})(^{62}_{59})$} & \multicolumn{2}{ |c| }{$+7.13(1.48)(^{1.59}_{1.54})(^{85}_{80})$} \\ [0.3ex]
\hline
Total & $-0.14(6)$ & $-0.23(^{11}_{9})$ & $-1.39(^{62}_{64})$ & $-2.26(^{1.12}_{0.93})$ & $-7.92(^{1.83}_{2.13})$ & $-12.86(^{5.75}_{4.15})$ & $-9.45(^{2.51}_{2.83})$ & $-15.35(^{6.98}_{5.17})$ \\ [0.3ex]
\hline
\end{tabular}}
\caption{$\Delta a_\mu^{\text{HVP,LO}}$ in units of $10^{-10}$. The uncertainties in the fifth row (FF) are related to $\Delta\Gamma_\rho$, $\Delta M_\rho$ and $\theta_{\omega\rho}$, respectively.}
\label{tab:Delta_amu}
\end{table}

Using the $\tau$ spectral functions measured by ALEPH~\cite{ALEPH:2005qgp}, Belle~\cite{Belle:2008xpe}, CLEO~\cite{CLEO:1999dln} and OPAL~\cite{OPAL:1998rrm}, we evaluate $a_\mu^{\text{HVP, LO}}[\pi\pi]$ using the window parameters in Eq. (\ref{eq:win_parameters}). These results are outlined in Tables \ref{HVP:tab4.4a} and \ref{HVP:tab4.4b} for $s\leq 1,\,2,\,3,\text{ and }3.125\,\text{ GeV}^2$, i.e., integrating Eq. (\ref{amu_dispersive}) with $\sigma_{\pi\pi(\gamma)}^0$ in Eq. (\ref{eq:pipi_cross_section}) from $s_{\text{thr}}=4m_\pi^2$ up to some given cut-off. In the aforementioned tables 2 and 3, the first uncertainty is connected to the systematic errors on the mass spectrum, and from the $\tau$-mass and $V_{ud}$ uncertainties; the second error is associated to $B_{\pi\pi^0}$ and $B_{e}$; and the third one is due to the IB corrections. The \textit{Mean} value in the tables corresponds to the weighted average from the different window contributions for each experiment, the first error is related to the experimental measurements, while the second one comes from the IB corrections.

\begin{figure}
    \centering
    \includegraphics[width=5.3cm]{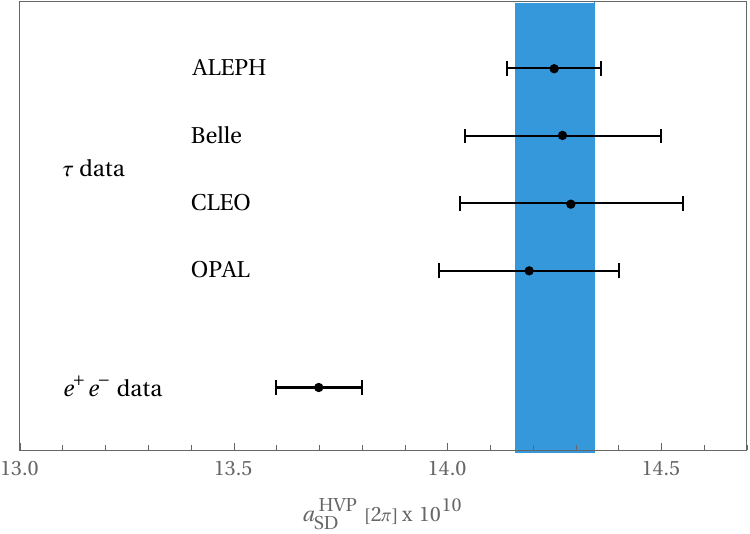}
    \includegraphics[width=5.2cm]{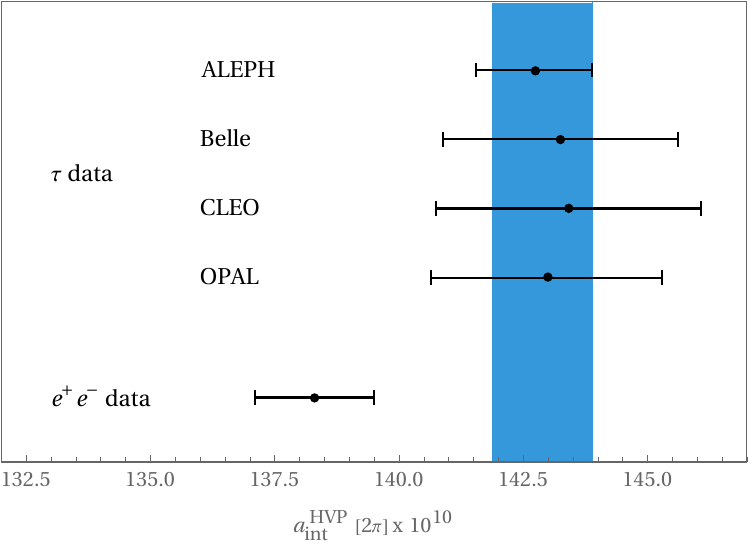}
    \includegraphics[width=5.3cm]{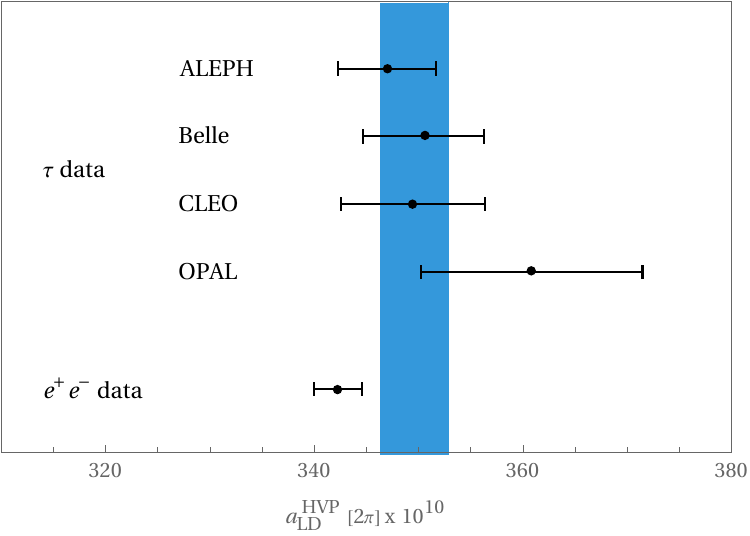}
    \caption{Windows quantities for HVP at $\mathcal{O}(p^4)$ for $2\pi$ below $1.0\,\text{GeV}$ using the parameters in Eq. (\ref{eq:win_parameters}). The blue region corresponds to the experimental average from $\tau$ data. The $e^+e^-$ number was taken from Ref. \cite{Colangelo:2022vok}. }
    \label{fig:ChPTOp4}
\end{figure}

\begin{figure}
    \centering
    \includegraphics[width=5.3cm]{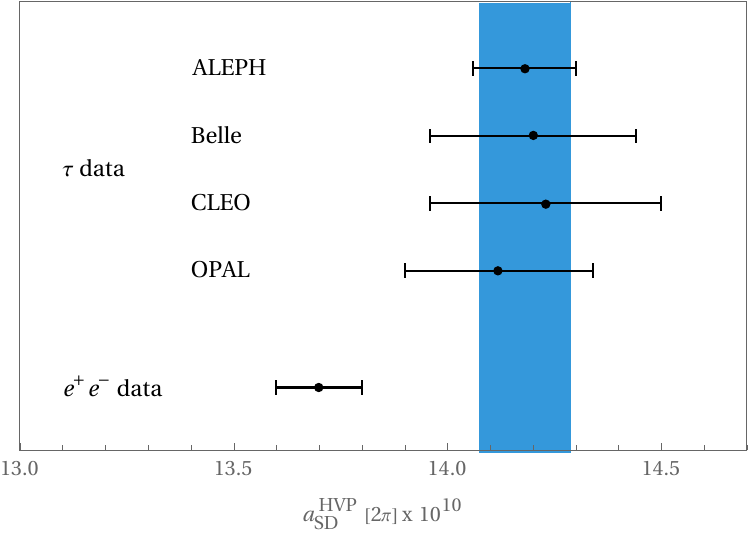}
    \includegraphics[width=5.2cm]{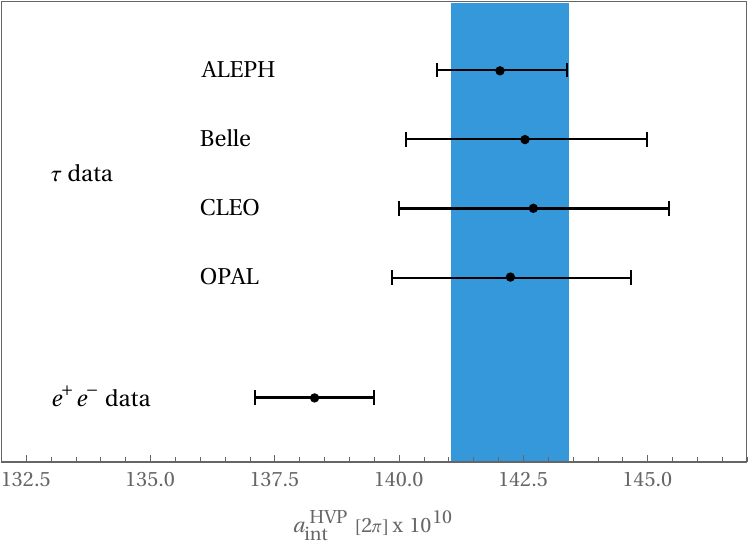}
    \includegraphics[width=5.3cm]{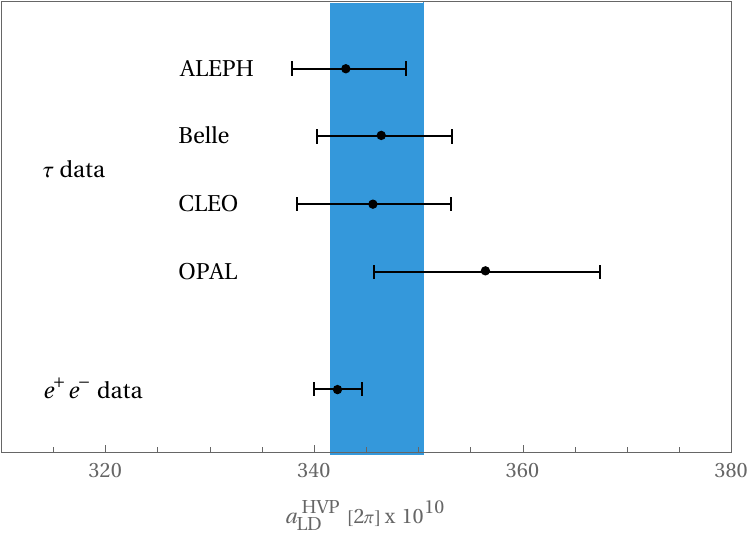}
    \caption{Analog to Fig. \ref{fig:ChPTOp4} but at $\mathcal{O}(p^6)$.}
    \label{fig:ChPTOp6}
\end{figure}

\begin{figure}
    \centering
    \includegraphics[width=8cm]{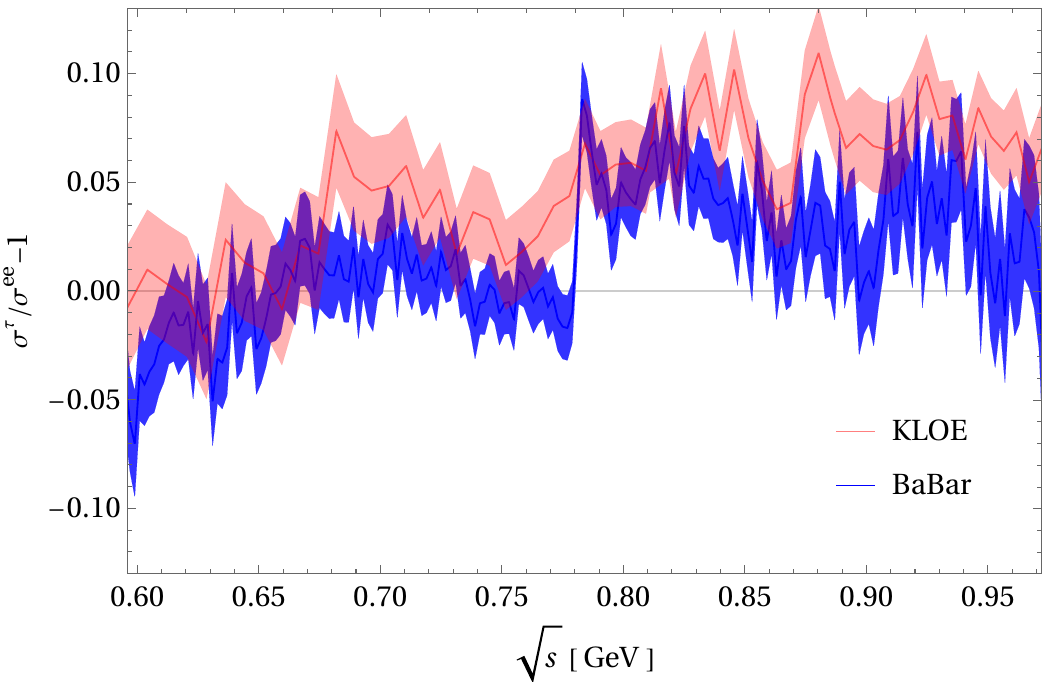}
    \includegraphics[width=8cm]{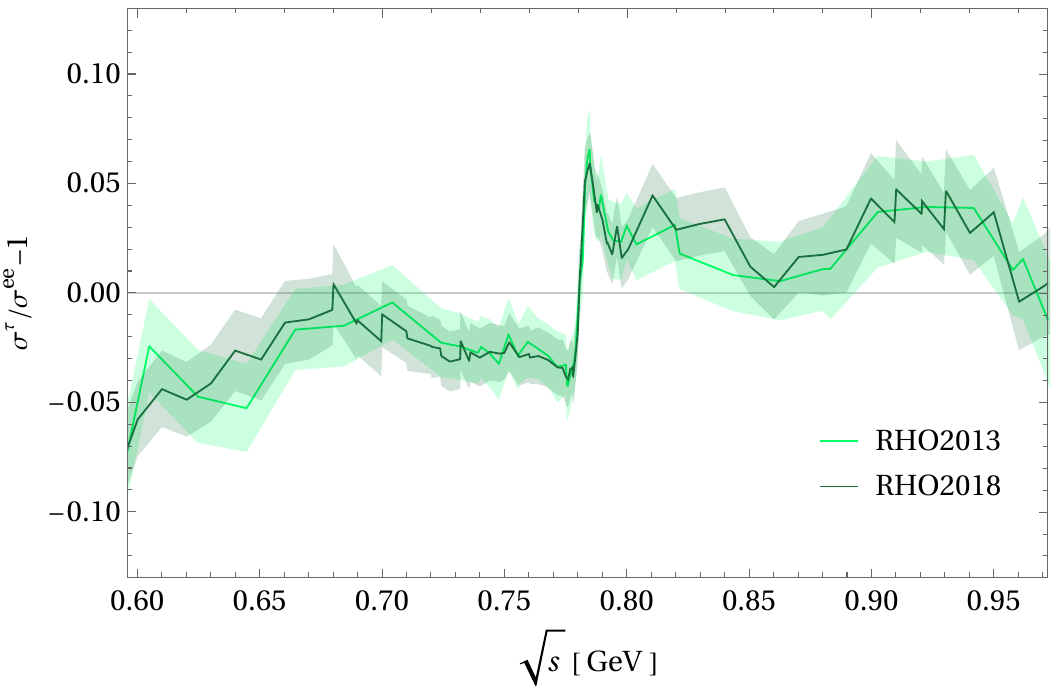}   
    \caption{Comparison between the $\tau$ (after IB corrections) and $e^+e^-\to\pi^+\pi^-$ spectral function using the ISR measurements from BABAR~\cite{BaBar:2012bdw} and KLOE~\cite{KLOE:2012anl} (left-hand) and the energy-scan measurements from CMD-3~\cite{CMD-3:2023alj} (right-hand). }
    \label{fig:KLEO_BABAR}
\end{figure}

An evaluation of $a_\mu^{\text{HVP}}$ in the windows in Euclidean time using $e^+e^-$ data was performed in Ref.~\cite{Colangelo:2022vok} using the parameters in Eq. (\ref{eq:win_parameters}), see Table~\ref{HVP:tab4.4c} below. A comparison between these window quantities for HVP in the $2\pi$ channel below $1\,\text{GeV}$ amounts to a discrepancy of $4.0\,\sigma$, $2.9\,\sigma$ and $1.8\,\sigma$ between the $\tau$ and $e^+e^-$ evaluations applying the $G_{\text{EM}}(s)$ correction at $\mathcal{O}(p^4)$ in R$\chi$T to the $\tau$ data for the $SD$, $int$ and $LD$ contributions, respectively. On the other hand, when we include the corrections at $\mathcal{O}(p^6)$, the difference between these two evaluations decreases to $3.3\,\sigma$, $2.3\,\sigma$ and $0.7\,\sigma$ for the $SD$, $int$ and $LD$ contributions, respectively. These results are depicted in Figs. \ref{fig:ChPTOp4} and \ref{fig:ChPTOp6}, where the blue band corresponds to the experimental average using $\tau$ data. Fig. \ref{fig:KLEO_BABAR} shows a zoomed comparison between $\tau$ (after IB corrections at $\mathcal{O}(p^6)$) and $e^+e^-\to\pi^+\pi^-$ spectral function using the ISR measurements from BABAR~\cite{BaBar:2012bdw} and KLOE~\cite{KLOE:2012anl} (left-hand panel) and the energy-scan measurements from CMD-3~\cite{CMD-3:2023alj} (right-hand panel). Colored bands correspond to the weighted average of the uncertainties coming from both sets of data in each figure. Although it may seem obvious than increased IB-corrections in the $\rho$ region will increase the compatibility between $\tau$ and CMD-3 $e^+e^-$ data, dedicated studies (like e.g. Ref.~\cite{Davier:2023cyp}) seem necessary to fully understand this issue (even more in the comparison with KLOE and BaBar).

\begin{table}[htbp]
\centering
\resizebox{16cm}{!}{\begin{tabular}{|c|c|c|c|c|}
\hline 
 \multicolumn{5}{ |c| }{$a_\mu^{\text{HVP,LO}}[\pi\pi,\tau]$} \\ [0.3ex]
\hline
 \multicolumn{5}{ |c| }{SD} \\ [0.3ex]
\hline
Experiment & $s\leq 1\,\mathrm{GeV}^2$ & $s\leq 2\,\mathrm{GeV}^2$ & $s\leq 3\,\mathrm{GeV}^2$ & $s\leq 3.125\,\mathrm{GeV}^2$\\ [0.3ex]
\hline
ALEPH & $14.25(4)(8)(6)$ & $15.46(2)(9)(6)$ & $15.52(3)(9)(6)$ & $15.52(3)(9)(6)$\\ [0.3ex]
Belle & $14.27(4)(22)(6)$ & $15.34(5)(24)(6)$ & $15.39(6)(24)(6)$ & $15.41(12)(24)(6)$\\ [0.3ex]
CLEO  & $14.29(6)(25)(6)$ & $15.41(6)(27)(6)$ & $15.45(6)(27)(6)$ & $15.46(6)(27)(6)$\\ [0.3ex]
OPAL  & $14.19(7)(19)(6)$ & $15.46(3)(21)(6)$ & $15.51(3)(21)(6)$ & $15.51(3)(21)(6)$\\ [0.3ex]
Mean & $14.25(7)(6)$%{ $[+4.3\sigma]$}
& $15.44(8)(6)$ & $15.50(8)(6)$ & $15.50(8)(6)$ \\ [0.3ex]
\hline
 \multicolumn{5}{ |c| }{Intermediate} \\ [0.3ex]
\hline
Experiment & $s\leq 1\,\mathrm{GeV}^2$ & $s\leq 2\,\mathrm{GeV}^2$ & $s\leq 3\,\mathrm{GeV}^2$ & $s\leq 3.125\,\mathrm{GeV}^2$\\ [0.3ex]
\hline
ALEPH & $142.73(61)(79)(^{61}_{63})$   & $149.02(50)(83)(^{61}_{63})$ & $149.18(48)(83)(^{61}_{63})$ & $149.18(48)(83)(^{61}_{63})$\\ [0.3ex]
Belle & $143.25(45)(2.24)(^{61}_{63})$   & $148.83(48)(2.33)(^{62}_{64})$ & $148.97(49)(2.33)(^{62}_{64})$ & $149.00(54)(2.33)(^{62}_{64})$\\ [0.3ex]
CLEO  & $143.41(60)(2.52)(^{61}_{63})$   & $149.23(61)(2.62)(^{62}_{64})$ & $149.34(61)(2.62)(^{62}_{64})$ & $149.37(61)(2.62)(^{62}_{64})$\\ [0.3ex]
OPAL  & $142.98(1.17)(1.92)(^{60}_{62})$ & $149.56(97)(2.01)(^{60}_{63})$ & $149.69(97)(2.01)(^{60}_{63})$ & $149.69(97)(2.01)(^{60}_{63})$\\ [0.3ex]
Mean & $142.89(81)(^{61}_{63})$%{ $[+3.2\sigma]$} 
& $149.09(80)(^{61}_{64})$ & $149.23(79)(^{61}_{64})$ & $149.24(79)(^{61}_{64})$ \\ [0.3ex]
\hline
 \multicolumn{5}{ |c| }{LD} \\ [0.3ex]
\hline
Experiment & $s\leq 1\,\mathrm{GeV}^2$ & $s\leq 2\,\mathrm{GeV}^2$ & $s\leq 3\,\mathrm{GeV}^2$ & $s\leq 3.125\,\mathrm{GeV}^2$\\ [0.3ex]
\hline
ALEPH & $347.03(4.12)(1.93)(^{1.27}_{1.47})$   & $350.79(4.08)(1.95)(^{1.27}_{1.47})$ & $350.83(4.08)(1.95)(^{1.27}_{1.47})$ & $350.83(4.08)(1.95)(^{1.27}_{1.47})$\\ [0.3ex]
Belle & $350.55(1.36)(5.49)(^{1.27}_{1.48})$   & $353.90(1.37)(5.54)(^{1.28}_{1.49})$ & $353.93(1.37)(5.55)(^{1.28}_{1.49})$ & $353.93(1.37)(5.55)(^{1.27}_{1.48})$\\ [0.3ex]
CLEO  & $349.49(2.88)(6.13)(^{1.27}_{1.47})$   & $353.00(2.88)(6.20)(^{1.27}_{1.47})$ & $353.02(2.88)(6.20)(^{1.27}_{1.48})$ & $353.03(2.88)(6.20)(^{1.27}_{1.47})$\\ [0.3ex]
OPAL  & $360.91(9.33)(4.85)(^{1.26}_{1.51})$ & $364.82(9.26)(4.90)(^{1.25}_{1.51})$ & $364.84(9.26)(4.90)(^{1.26}_{1.51})$ & $364.84(9.26)(4.90)(^{1.26}_{1.51})$\\ [0.3ex]
Mean & $349.65(3.01)(^{1.27}_{1.48})$%{ $[+2.1\sigma]$}
& $353.25(3.01)(^{1.27}_{1.48})$ & $353.28(3.01)(^{1.27}_{1.49})$ & $353.28(3.01)(^{1.27}_{1.48})$ \\ [0.3ex]
\hline
\end{tabular}}
\caption{IB-corrected $a_\mu^{\text{HVP,LO}}[\pi\pi,\tau]$ in units of $10^{-10}$ at $\mathcal{O}(p^4)$ in R$\chi T$ using the experimental measurements from the ALEPH~\cite{ALEPH:2005qgp}, Belle~\cite{Belle:2008xpe}, CLEO~\cite{CLEO:1999dln} and OPAL~\cite{OPAL:1998rrm} Colls. The first error is related to the systematic uncertainties on the mass spectrum and also includes contributions from the $\tau$-mass and $V_{ud}$ uncertainties. The second error arises from $B_{\pi\pi^0}$ and $B_e$, and the third error comes from the isospin-breaking corrections. The uncertainties in the mean value correspond to the experiment and to the IB corrections, respectively.}
\label{HVP:tab4.4a}
\end{table}
%%%%%%%%%%%%%%%%%%%%%%%%%%%%%%%%%%%%%%%%%%%%%%%%%%%%%%%%%%%%%%
\begin{table}[htbp]
\centering
\resizebox{16cm}{!}{\begin{tabular}{|c|c|c|c|c|}
\hline
 \multicolumn{5}{ |c| }{$a_\mu^{\text{HVP,LO}}[\pi\pi,\tau]$} \\ [0.3ex]
\hline
 \multicolumn{5}{ |c| }{SD} \\ [0.3ex]
\hline
Experiment & $s\leq 1\,\mathrm{GeV}^2$ & $s\leq 2\,\mathrm{GeV}$ & $s\leq 3\,\mathrm{GeV}$ & $s\leq 3.125\,\mathrm{GeV}^2$\\ [0.3ex]
\hline
ALEPH  & $14.18(4)(8)(8)$ & $15.37(2)(9)(^{11}_{9})$ & $15.42(2)(9)(^{11}_{9})$ & $15.42(3)(9)(^{11}_{9})$\\ [0.3ex]
Belle & $14.20(4)(22)(8)$ & $15.25(5)(24)(^{10}_{8})$ & $15.30(5)(24)(^{10}_{8})$ & $15.31(12)(24)(^{10}_{8})$\\ [0.3ex]
CLEO  & $14.23(6)(25)(8)$ & $15.32(6)(27)(^{10}_{8})$ & $15.36(6)(27)(^{10}_{9})$ & $15.37(6)(27)(^{10}_{9})$\\ [0.3ex]
OPAL  & $14.12(7)(19)(8)$ & $15.37(3)(21)(^{11}_{9})$ & $15.41(3)(21)(^{11}_{9})$ & $15.41(3)(21)(^{11}_{9})$\\ [0.3ex]
Mean & $14.18(7)(8)$%{ $[+3.6\sigma]$}
& $15.35(8)(^{11}_{9})$ & $15.40(8)(^{11}_{9})$ & $15.41(8)(^{11}_{9})$ \\ [0.3ex]
\hline
 \multicolumn{5}{ |c| }{Intermediate} \\ [0.3ex]
\hline
Experiment & $s\leq 1\,\mathrm{GeV}^2$ & $s\leq 2\,\mathrm{GeV}^2$ & $s\leq 3\,\mathrm{GeV}^2$ & $s\leq 3.125\,\mathrm{GeV}^2$\\ [0.3ex]
\hline
ALEPH & $142.04(60)(79)(^{94}_{83})$   & $148.21(50)(83)(^{1.04}_{89})$ & $148.35(48)(83)(^{1.04}_{89})$ & $148.35(48)(83)(^{1.04}_{89})$\\ [0.3ex]
Belle & $142.55(45)(2.23)(^{94}_{83})$   & $148.03(48)(2.32)(^{1.04}_{90})$ & $148.16(48)(2.32)(^{1.05}_{90})$ & $148.19(54)(2.32)(^{1.05}_{90})$\\ [0.3ex]
CLEO  & $142.71(60)(2.50)(^{94}_{83})$   & $148.41(60)(2.60)(^{1.03}_{89})$ & $148.52(60)(2.61)(^{1.03}_{89})$ & $148.55(60)(2.61)(^{1.03}_{89})$\\ [0.3ex]
OPAL  & $142.24(1.16)(1.91)(^{98}_{85})$ & $148.68(95)(2.00)(^{1.08}_{92})$ & $148.80(95)(2.00)(^{1.09}_{92})$ & $148.80(95)(2.00)(^{1.09}_{92})$\\ [0.3ex]
Mean & $142.19(80)(^{95}_{84})$%{ $[+2.6\sigma]$}
& $148.26(79)(^{1.05}_{0.90})$ & $148.40(79)(^{1.05}_{90})$ & $148.41(79)(^{1.05}_{0.90})$ \\ [0.3ex]
\hline
 \multicolumn{5}{ |c| }{LD} \\ [0.3ex]
\hline
Experiment & $s\leq 1\,\mathrm{GeV}^2$ & $s\leq 2\,\mathrm{GeV}^2$ & $s\leq 3\,\mathrm{GeV}^2$ & $s\leq 3.125\,\mathrm{GeV}^2$\\ [0.3ex]
\hline
ALEPH & $343.09(3.99)(1.91)(^{3.65}_{2.78})$   & $346.78(3.95)(1.93)(^{3.71}_{2.83})$ & $346.81(3.95)(1.93)(^{3.71}_{2.83})$ & $346.81(3.95)(1.93)(^{3.71}_{2.83})$\\ [0.3ex]
Belle & $346.53(1.34)(5.43)(^{3.75}_{2.86})$   & $349.82(1.35)(5.48)(^{3.80}_{2.90})$ & $349.84(1.35)(5.48)(^{3.80}_{2.90})$ & $349.84(1.36)(5.48)(^{3.80}_{2.90})$\\ [0.3ex]
CLEO  & $345.55(2.77)(6.06)(^{3.67}_{2.80})$   & $348.99(2.77)(6.12)(^{3.72}_{2.85})$ & $349.01(2.77)(6.12)(^{3.72}_{2.85})$ & $349.02(2.77)(6.13)(^{3.72}_{2.85})$\\ [0.3ex]
OPAL  & $356.42(8.99)(4.79)(^{4.21}_{3.18})$ & $360.25(8.92)(4.84)(^{4.27}_{3.22})$ & $360.28(8.92)(4.84)(^{4.27}_{3.22})$ & $360.28(8.92)(4.84)(^{4.27}_{3.22})$\\ [0.3ex]
Mean & $345.65(2.95)(^{3.82}_{2.91})$%{ $[+0.9\sigma]$}
& $349.17(2.95)(^{3.88}_{2.95})$ & $349.19(2.95)(^{3.88}_{2.95})$ & $349.20(2.95)(^{3.88}_{2.95})$ \\ [0.3ex]
\hline
\end{tabular}}
\caption{Same as Table \ref{HVP:tab4.4a}, but the $\mathcal{O}(p^6)$ contributions to $G_{\text{EM}}(s)$ in Ref.~\cite{Miranda:2020wdg} have been applied to the $\tau$ data.}
\label{HVP:tab4.4b}
\end{table}
%%%%%%%%%%%%%%%%%%%%%%%%%%%%%%%%%%%%%%%%%%%%%%%%%%%%%%%%%%%%%%
%%%%%%%%%%%%%%%%%%%%%%%%%%%%%%%%%%%%%%%%%%%%%%%%%%%%%%%%%%%%%%%%
%%%%%%%%%%%%%%%%%%%%%%%%%%%%%%%%%%%%%%%%%%%%%%%%%%%%%%%%%%%%%%
%%%%%%%%%%%%%%%%%%%%%%%%%%%%%%%%%%%%%%%%%%%%%%%%%%%%%%%%%%%%%%

A direct comparison between $a_{\mu}^{\text{HVP, LO}}[\pi\pi,\tau]$ and the lattice results is not possible. For that endeavour, it is necessary to supplement the $2\pi$ evaluation with the remaining contributions from all other channels accounting for the hadronic cross-section. To illustrate the impact of this contribution in $a_{\mu}^{\text{HVP, LO}}$, we follow two approaches~\footnote{In this exploratory study correlations
between different energy ranges in the two pion channel and also between the two
pion and other channels are neglected. We plan to improve this within the joint effort of the muon $g-2$ theory initiative, https://muon-gm2-theory.illinois.edu/.}. Firstly, using the values reported in Table 1 of Ref.~\cite{Colangelo:2022vok} we subtract the contribution from the $2\pi$ channel below $1.0\,\text{GeV}$ (we represent this procedure with '$<1$ GeV') and replace it by the corresponding mean value in Tables \ref{HVP:tab4.4a} and \ref{HVP:tab4.4b}. This way, we get
\begin{equation}
a^{SD}_{\mu}=69.0(5)\times 10^{-10},\quad a^{int}_{\mu}=234.0(^{1.2}_{1.3})\times 10^{-10},\quad a^{LD}_{\mu}=402.5(^{3.3}_{3.4})\times 10^{-10},
\end{equation}
at $\mathcal{O}(p^4)$, and 
\begin{equation}
a^{SD}_{\mu}=68.9(5)\times 10^{-10},\quad a^{int}_{\mu}=233.3(1.4)\times 10^{-10},\quad a^{LD}_{\mu}=398.5(^{4.9}_{4.2})\times 10^{-10},
\end{equation}
at $\mathcal{O}(p^6)$.

Secondly, we estimate the $2\pi$ full contribution using the ratios $a_\mu^I/a_\mu^\text{HVP, LO}$, where $I$ stands for $SD$, $int$ and $LD$, from the corresponding window quantities in Ref.~\cite{Colangelo:2022vok} and the overall weighted-average evaluation of $a_\mu^\text{HVP, LO}[\pi\pi,e^+e^-]=505.1(1.7)\times 10^{-10}$ in Refs.~\cite{Davier:2019can,Keshavarzi:2019abf}. However, as can be easily computed from Tables \ref{HVP:tab4.4a} and \ref{HVP:tab4.4b}, these ratios are not the same between the second and the last column. This effect is mainly due to the weight functions $\tilde{\Theta}(s)$ in Eq. (\ref{eq:weight_function})~\footnote{We gratefully thank Michel Davier, Bogdan Malaescu and Zhiqing Zhang for pointing out an inconsistency in our earlier procedure, that we have now amended.}, so, in order to take this  into account, we use the difference between the ratio from the second and third column to correct the ratios from $e^+e^-$ data. 
Then we subtract this value from the total contribution and replace it by our results~\footnote{Additionally, we have included a $50\%$ uncertainty due to the difference between the ratios in the second and the last column to the final result.}. Finally, we get
\begin{equation}
a^{SD}_{\mu}=69.0(7)\times 10^{-10},\quad a^{int}_{\mu}=234.2(2.0)\times 10^{-10},\quad a^{LD}_{\mu}=402.6(^{3.8}_{3.9})\times 10^{-10},
\end{equation}
at $\mathcal{O}(p^4)$, and 
\begin{equation}
a^{SD}_{\mu}=68.9(7)\times 10^{-10},\quad a^{int}_{\mu}=233.4(2.1)\times 10^{-10},\quad a^{LD}_{\mu}=398.5(^{5.3}_{4.6})\times 10^{-10},
\end{equation}
at $\mathcal{O}(p^6)$. All these results are reasonably consistent with each other.

\begin{figure}
    \centering
    \includegraphics[width=9cm]{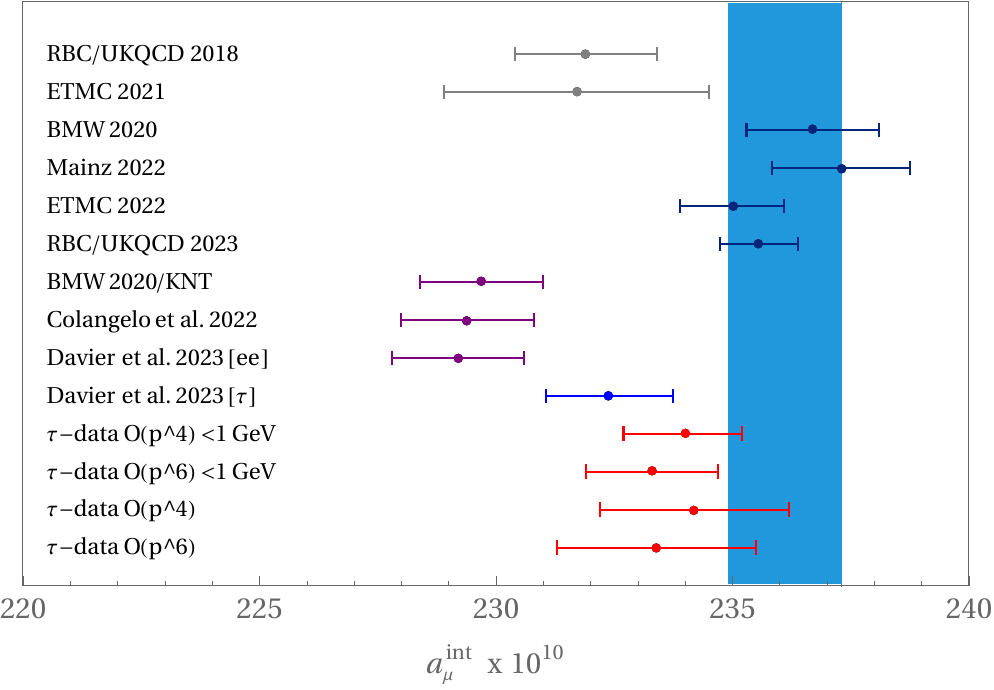}
    \caption{Comparison of different evaluations of the total intermediate window contribution to $a_{\mu}^{\text{HVP, LO}}$. The blue band corresponds to the weighted average of the lattice results excluding RBC/UKQCD 2018~\cite{RBC:2018dos} and ETMC 2021~\cite{Giusti:2021dvd}.}
    \label{fig:lattice_results}
\end{figure}

We summarize these outcomes in Table \ref{HVP:tab4.4c} along with the lattice results~\cite{RBC:2018dos,Giusti:2021dvd,Borsanyi:2020mff,Ce:2022kxy,ExtendedTwistedMass:2022jpw,Blum:2023qou} and other $e^+e^-$ data-driven evaluations~\cite{Aoyama:2020ynm,Colangelo:2022vok}. These numbers are depicted in Fig. \ref{fig:lattice_results} for the intermediate window, where  the blue band represents the weighted average of the lattice results, $a_\mu^{int}=235.8(6)\cdot 10^{-10}$, excluding those from RBC/UKQCD 2018~\cite{RBC:2018dos} and ETMC 2021~\cite{Giusti:2021dvd} collaborations. The contributions of the intermediate window using $\tau$ data are slightly closer to the results from lattice QCD than to the $e^+e^-$ values. Therefore, the $\sim 4.3\sigma$ discrepancy between the $e^+e^-$ data-driven and lattice evaluations is reduced to $\sim 1.5\sigma$ when $\tau$ data is used for the $2\pi$ channel. On the other hand, there is only one lattice result for the short-distance window~\cite{ExtendedTwistedMass:2022jpw} which seems to be in agreement with both data-driven HVP evaluations.

\begin{table}[htbp]
\centering
\resizebox{11cm}{!}{\begin{tabular}{|c|c|c|c|c|}
\hline
 \multicolumn{5}{ |c| }{$a_\mu^{\text{HVP,LO}}$} \\ [0.3ex]
\hline
 & SD & int & LD & Total\\ [0.3ex]
\hline
$\tau$-data $\mathcal{O}(p^4)$ $\leq 1\text{ GeV}$ & $69.0(5)$ & $234.0(^{1.2}_{1.3})$ & $402.5(^{3.3}_{3.4})$ & $705.5(^{5.0}_{5.2})$ \\ [0.3ex]
$\tau$-data $\mathcal{O}(p^6)$ $\leq 1\text{ GeV}$ & $68.9(5)$ & $233.3(1.4)$ & $398.5(^{4.9}_{4.2})$ & $700.7(^{6.8}_{6.1})$ \\ [0.3ex]
\hline
$\tau$-data $\mathcal{O}(p^4)$ & $69.0(7)$ & $234.2(2.0)$ & $402.6(^{3.8}_{3.9})$ & $705.8(^{6.5}_{6.6})$\\ [0.3ex]
$\tau$-data $\mathcal{O}(p^6)$ & $68.9(7)$ & $233.4(2.1)$ & $398.5(^{5.3}_{4.6})$ & $700.8(^{8.1}_{7.4})$\\ [0.3ex]
\hline
RBC/UKQCD 2018~\cite{RBC:2018dos}  & $-$ & $231.9(1.5)$ & $-$ & $715.4(18.7)$\\ [0.3ex]
ETMC 2021~\cite{Giusti:2021dvd}  & $-$ & $231.7(2.8)$ & $-$ & $-$ \\ [0.3ex]
BMW 2020~\cite{Borsanyi:2020mff}  & $-$ & $236.7(1.4)$ & $-$ & $707.5(5.5)$\\ [0.3ex]
Mainz/CLS 2022~\cite{Ce:2022kxy}  & $-$ & $237.30(1.46)$ & $-$ & $-$\\ [0.3ex]
ETMC 2022~\cite{ExtendedTwistedMass:2022jpw}  & $69.33(29)$ & $235.0(1.1)$ & $-$ & $-$ \\ [0.3ex]
RBC/UKQCD 2023~\cite{Blum:2023qou}  & $-$ & $235.56(82)$ & $-$ & $-$\\ [0.3ex]
\hline
WP~\cite{Aoyama:2020ynm} & $-$ & $-$ & $-$ & $693.1(4.0)$\\ [0.3ex]
BMW 2020/KNT~\cite{Keshavarzi:2018mgv,Borsanyi:2020mff}  & $-$ & $229.7(1.3)$ & $-$ & $-$ \\ [0.3ex]
Colangelo et al. 2022~\cite{Colangelo:2022vok} & $68.4(5)$ & $229.4(1.4)$ & $395.1(2.4)$ & $693.0(3.9)$\\ [0.3ex]
Davier et al. 2023 [$e^+e^-$]~\cite{Davier:2023cyp} & $-$ & $229.2(1.4)$ & $-$ & $694.0(4.0)$\\ [0.3ex]
Davier et al. 2023 [$\tau$]~\cite{Davier:2023fpl} & $-$ & $232.4(1.3)$ & $-$ & $-$\\ [0.3ex]
\hline
\end{tabular}}
\caption{Window quantities for $a_\mu^{\text{HVP,LO}}$ in units of $10^{-10}$. The first two rows correspond to the $\tau$ evaluation in the first approach, while rows 3 and 4 are the evaluations in the second one. %\textbf{(I would add the uncertainties in the total of the first four rows)}
The rows 5-10 are the lattice results~\cite{RBC:2018dos,Giusti:2021dvd,Borsanyi:2020mff,Ce:2022kxy,ExtendedTwistedMass:2022jpw,Blum:2023qou}. The last rows are the evaluations obtained using $e^+e^-$~\cite{Aoyama:2020ynm,Keshavarzi:2018mgv,Borsanyi:2020mff,Colangelo:2022vok,Davier:2023cyp} and $\tau$ data~\cite{Davier:2023fpl}. See Fig. 11 in Ref.~\cite{Blum:2023qou} for more details.}
\label{HVP:tab4.4c}
\end{table}

\section{Conclusions}\label{sec:Concl}
While the BNL and FNAL measurements of $a_\mu$ agree nicely within errors, the situation is not that clear for its SM prediction's counterpart. On the one hand, data-driven methods based on $e^+e^-\to$hadrons data have to deal with the tensions between experiments (particularly among BaBar and KLOE, and now with CMD-3), which makes the  computation of the uncertainty in Eq.~(\ref{eq:amuSM}) a non-trivial task, as will be its update. On the other side, there is still only one lattice QCD evaluation (BMWc Coll.) of $a_\mu^{\text{HVP}}$ with competitive uncertainties, that lies between $a_\mu^{\mathrm{Exp}}$ and its SM data-driven prediction. However, the most recent Mainz/CLS, ETMC, and RBC/UKQCD results have similar errors to BMWc in the intermediate window, where all of them agree remarkably. It is long-known that alternative data-driven evaluations are possible, utilizing this time semileptonic $\tau$-decay data (and isospin-breaking corrections, with attached model-dependent uncertainty), as we have done here.

In this context, we have applied the study from Ref.~\cite{Blum:2023qou}, that computed window quantities in Euclidean time for data-driven evaluations of $a_\mu^{\text{HVP}}$ using $e^+e^-\to$ hadrons data, to the semileptonic $\tau$ decays case (focusing on the dominant two-pion contribution). Our main results are collected in Table \ref{HVP:tab4.4c} and show that $\tau$-based results are compatible with the lattice evaluations in the intermediate window, being the $e^+e^-$-based values in tension with both of them. This difference is the main cause for the larger discrepancy of the latter with $a_\mu^{\mathrm{Exp}}$ and should be further scrutinized. Supplemented by the relevant IB corrections computed on the lattice \cite{Bruno:2018ono}, the results in this work could be used together with lattice QCD to obtain
an alternative data-based determination from $\tau$ decays (and lattice QCD) which can be helpful in solving the present puzzle.

\section*{Acknowledgements}
We have benefited from discussions with Rafel Escribano on this topic. We are indebted to Michel Davier, Gabriel López Castro, Bogdan Malaescu and Zhiqing Zhang for extremely useful discussions and detailed comparisons of our mutual results. It is always instructive and enlightening to learn something from Vincenzo Cirigliano. We thank the anonymous referees for their very useful reports, which helped us to improve this letter. The work of P. M. has been supported by the European Union’s Horizon 2020 Research
and Innovation Programme under grant 824093 (H2020-INFRAIA- 2018-1), the Ministerio de Ciencia e Innovación under grant PID2020-112965GB-I00, and by the Secretaria
d’Universitats i Recerca del Departament d’Empresa i Coneixement de la Generalitat de
Catalunya under grant 2021 SGR 00649. IFAE is partially funded by the CERCA program
of the Generalitat de Catalunya. J. A. M. is also supported by MICINN with funding from
European Union NextGenerationEU (PRTR-C17.I1) and by Generalitat de Catalunya. 
P.~R. thanks partial funding from Conacyt and Cátedras Marcos Moshinsky 2020 (Fundación Marcos Moshinsky).

\end{document}